\documentclass[fleqn,10pt]{wlscirep}
\usepackage[utf8]{inputenc}
\usepackage[T1]{fontenc}
\usepackage{xcolor,colortbl}
\usepackage{siunitx}
\usepackage{cite} % for multiple reference citation

\definecolor{Gray}{gray}{0.75}
\newcolumntype{a}{>{\columncolor{Gray}}c}

\title{Automated Detection of Cribriform Growth Patterns in Prostate Histology Images}

\author[1,*]{Pierre Ambrosini}
\author[2]{Eva Hollemans}
\author[2]{Charlotte F. Kweldam}
\author[2]{Geert J. L. H. van Leenders}
\author[1,+]{Sjoerd Stallinga}
\author[1,3,+]{Frans Vos}
\affil[1]{Department of Imaging Physics, Faculty of Applied Sciences, Delft University of Technology, Delft, The Netherlands}
\affil[2]{Department of Pathology, Erasmus University Medical Center, Rotterdam, The Netherlands}
\affil[3]{Department of Radiology, Academic Medical Center, Amsterdam, The Netherlands}

\affil[*]{p.ambrosini@tudelft.nl}
\affil[+]{These authors contributed equally to this work}


\begin{abstract}
		Cribriform growth patterns in prostate carcinoma are associated with poor prognosis. We aimed to introduce a deep learning method to detect such patterns automatically.
		To do so, convolutional neural network was trained to detect cribriform growth patterns on 128 prostate needle biopsies. Ensemble learning taking into account other tumor growth patterns during training was used to cope with heterogeneous and limited tumor tissue occurrences.
		ROC and FROC analyses were applied to assess network performance regarding detection of biopsies harboring cribriform growth pattern.
		The ROC analysis yielded a mean area under the curve up to 0.81.
		FROC analysis demonstrated a sensitivity of 0.9 for regions larger than \SI{0.0150}{\milli\metre\squared} with on average 7.5 false positives.
		To benchmark method performance for intra-observer annotation variability, false positive and negative detections were re-evaluated by the pathologists.
		Pathologists considered 9\% of the false positive regions as cribriform, and 11\% as possibly cribriform; 44\% of the false negative regions were not annotated as cribriform.
		As a final experiment, the network was also applied on a dataset of 60 biopsy regions annotated by 23 pathologists.
		With the cut-off reaching highest sensitivity, all images annotated as cribriform by at least $7/23$ of the pathologists, were all detected as cribriform by the network and $9/60$ of the images were detected as cribriform whereas no pathologist labelled them as such.
		In conclusion, the proposed deep learning method has high sensitivity for detecting cribriform growth patterns at the expense of a limited number of false positives.
		It can detect cribriform regions that are labelled as such by at least a minority of pathologists. Therefore, it could assist clinical decision making by suggesting suspicious regions.
\end{abstract}
\begin{document}

\thispagestyle{empty}
\noindent
\textbf{Manuscript published in Scientific Reports, 10, 14904, 2020} \\
\textbf{DOI:10.1038/s41598-020-71942-7} \\
\textbf{The final version of the paper is available at:} \\
\textbf{\url{https://doi.org/10.1038/s41598-020-71942-7}} \\

\flushbottom
\maketitle

\thispagestyle{empty}

\section*{Introduction}

\begin{figure} % h
				\centering
				\includegraphics[width=0.9\linewidth]{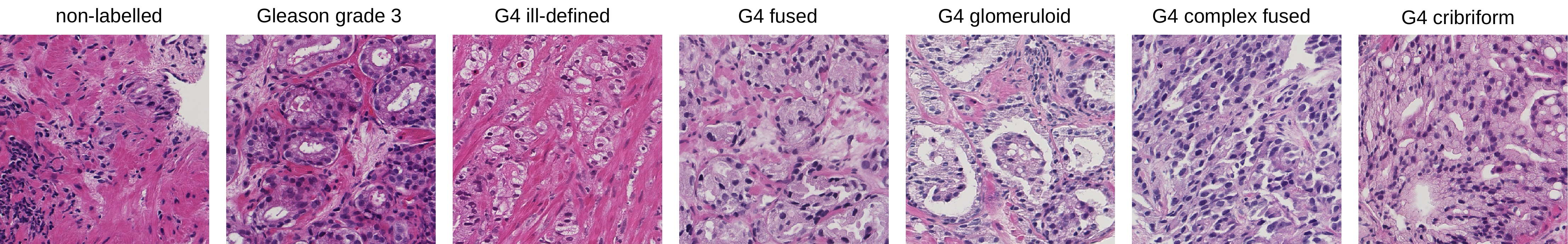}
				\\ \vspace{5mm}
				\includegraphics[width=0.8\linewidth]{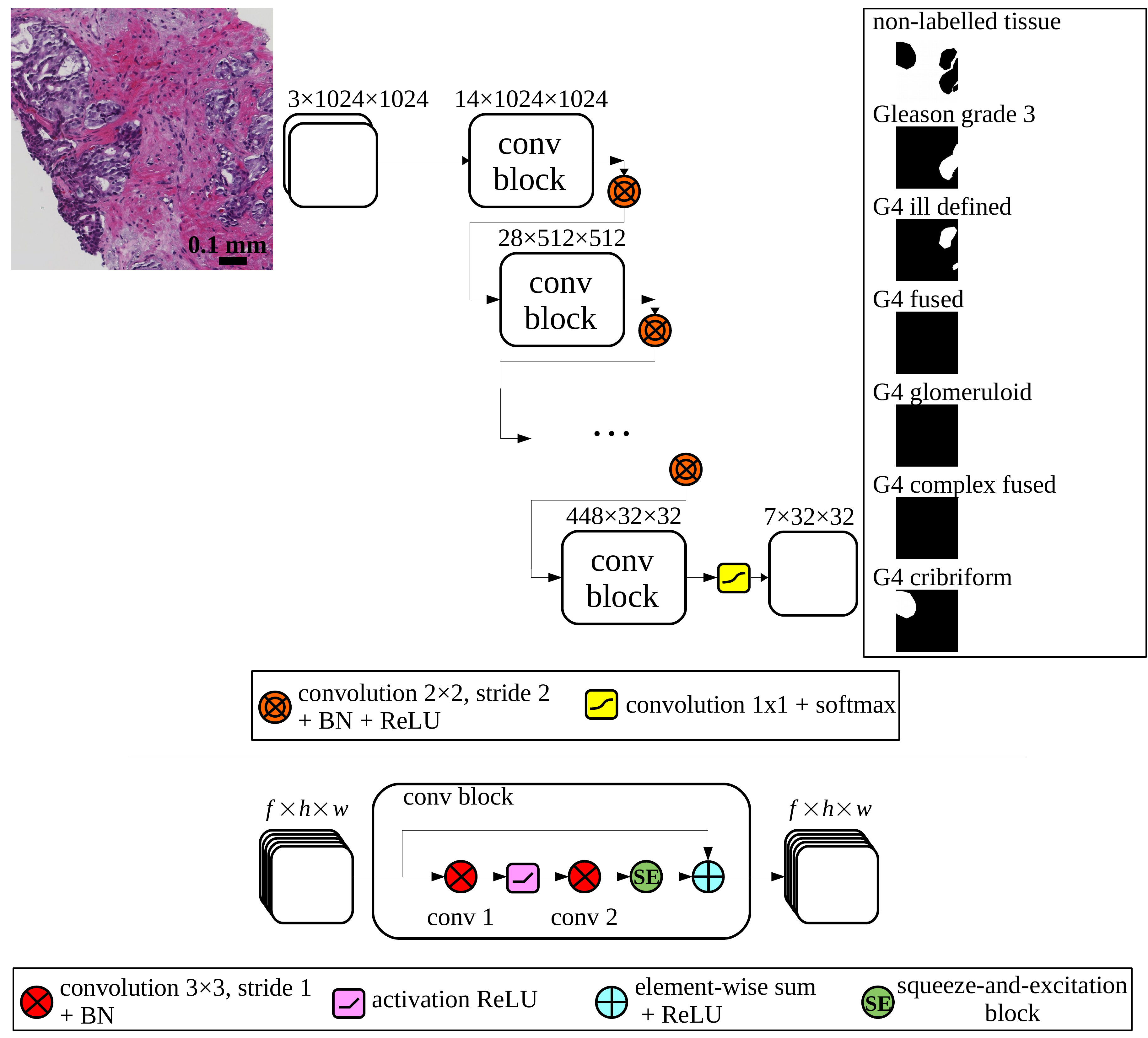}
				\caption{(top) Examples of different biopsy tissues with glands classified as Gleason grade 3 or 4.\\
				(bottom) Deep convolutional neural network architecture: a biopsy patch image is fed into the network and the output after a softmax normalization is composed of 7 segmentation maps 32 times smaller than the input. The network is composed of 6 consecutive \emph{conv blocks}. A \emph{conv block} consists of $2$ consecutive convolutions of input features ($f$ feature maps of height $h$ and width $w$) with a squeeze-and-excitation block and a residual connection. Downsampling is done using strided convolutions with a kernel 2x2, after which batch normalization (BN) and an activation rectifier linear unit (ReLU) are applied.}
				\label{fig:unet_classification3}
			\end{figure}

		Prostate cancer is one of the most common cancer types in men: about one man in 9 is diagnosed with prostate cancer in his lifetime \cite{cancerorg}. Histological image analysis of biopsy specimens is generally considered the reference standard for detection and grading of prostate cancer. The Gleason grading system is often used in practice to evaluate the severity of prostate cancer. The  Gleason system distinguishes five basic architectural growth patterns, numbered Gleason grade G1 to G5. Presently, combinations of prevalent growth patterns are usually considered which is reflected in the Gleason Score and Grade Group (Table~\ref{table:tableGleason}). However, in spite of such updates the  current system is still associated with high inter-observer variability \cite{chen2016evolving}. For example, the classification between Grade Group 2 and 3 (Gleason Score $3+4=7$ and $4+3=7$) is often subject to disagreement among pathologists. The classification between these two Grade Groups is highly relevant since it influences therapeutic decision-making.
		Actually, each individual Gleason grade is a collection of different growth patterns (Fig.~\ref{fig:unet_classification3}). Particularly, Gleason grade 4 comprises glands forming cribriform, glomeruloid, ill-defined, fused, and complex fused growth patterns \cite{epstein20162014,kweldam2019grading}. Unfortunately, the disagreement among pathologists is also relatively high regarding the sub-type classification.
	
		In a recent inter-observer study, high consensus, i.e. 80\% agreement among 23 pathologists, was reported on 23\% of the cribriform cases, but rarely on fused and never on ill-defined patterns \cite{kweldam2016gleason}. At the same time it was found that presence of cribriform growth patterns in prostate cancer imply a poor prognosis \cite{kweldam2015cribriform,kweldam2016disease,hollemans2019large}. The cribriform growth pattern could therefore be an important prognostic marker and its detection might add valuable information on top of the Gleason grading system.
		
		Many clinical decisions have to be made during the treatment of prostate cancer patients, often by multidisciplinary teams or tumor boards. These decisions are complex due to the increasing number of available parameters e.g. from radiological imaging, pathology and genomics. There is a clinical need for technology that enables objective, reproducible quantification of imaging features. Specifically, automatic qualification of the biopsies regarding the Gleason grade and biomarkers derived from automated detection of cribriform glands would add objective parameters in a clinical decision support algorithm. Such automated detection tools can also bring visualization support to aid clinical decision-making.
		
		We propose a method to automatically detect cribriform glands in prostate biopsy images. As the annotation of cribriform glands is subject to intra- and inter-observer variability, erroneous cases will be re-evaluated by the original annotators and the algorithm's performance will be compared against the assessments of a large group of pathologists.

		Engineered feature based machine learning approaches were used to identify stroma, benign and cancerous tissue for radical prostatectomy tissue slides \cite{gertych2015machine}. Automatic Gleason grading were proposed as well using multi-expert annotations and multi-scale features based methods \cite{nir2018automatic}. Additionally, deep learning methods have proven useful in digital pathology for various tasks such as detection and segmentation of glands, epithelium, stroma, cell nuclei and mitosis \cite{litjens2017survey}. This is relevant as such tissue segmentations can be a first step to a more detailed characterization.
		More recently, a variety of CNN-based methods were also used to classify prostate cancer tissue. These approaches differed regarding the type of histological images: Tissue Micro-Arrays (TMAs) \cite{arvaniti2018automated} and Whole Slide Images (WSIs) \cite{ing2018semantic,nagpal2018development} acquired after radical prostatectomy versus WSIs obtained from prostate needle biopsies.
		Biopsy interpretation is challenging, though, due to the narrow tissue width since typically a needle diameter of around 1~mm is used.
		Importantly, assessment of needle biopsies can have impact on management of individual patients.
		Previously, segmentation and classification methods for automatic processing of prostate needle biopsies were developed to detect malignant tissue \cite{litjens2016deep,campanella2019clinical}, as well as for partial \cite{lucas2019deep,li2019attention} and full Grade Group classification \cite{bulten2020automated,strom2020artificial}.
		Automated detection of cribriform growth patterns in prostate biopsy tissue has, to the best of our knowledge, not yet been studied. Indeed, Gertych et al. \cite{gertych2019convolutional} proposed a CNN combined with a soft-voting method to automatically distinguish four growth patterns including the cribriform growth pattern, but this was applied to lung tissue samples only. Moreover, it was stated that the method had only moderate recognition performance (F1-score=0.61) with regards to the cribriform growth pattern.
			
			In order to assist pathologists and support clinical decision making, we aim to introduce a method for automatic detection of cribriform growth patterns.
			In summary, this paper presents the following contributions:
			\begin{itemize}
				\item Cribriform growth patterns are automatically detected and segmented from tissue slides obtained from prostate needle biopsies.
				\item Annotations of erroneous cases are re-considered to account for intra-observer variability.
				\item Algorithm performance is compared against assessments by a large group of pathologists. 
			\end{itemize}
			
			\begin{table}[ht]
				\centering
				\begin{tabular}{|a|c|c|c|c|c|c|c|c|c|}
					\hline
					\textbf{Gleason grade} & 1,2 or 3 & 3 + 4 & 4 + 3 & 4 + 4 & 3 + 5 & 5 + 3 & 4 + 5 & 5 + 4 & 5 + 5 \\
					\hline
					\textbf{Gleason Score} & $\leq 6$ & 7 & 7 & 8 & 8 & 8 & 9 & 9 & 10 \\
					\hline
					\textbf{Grade Group} & 1 & 2 & 3 & 4 & 4 & 4 & 5 & 5 & 5 \\
					\hline
				\end{tabular}
				\caption{The Gleason grading system. A prostate specimen is classified in Gleason grade 1 to 5 \cite{chen2016evolving}. The Gleason Score of a tissue is the sum of the primary Gleason grade and the secondary Gleason grade (in terms of predominance). In order to differentiate tissues with Gleason Score $7=3+4$ and $7=4+3$, Grade Group classification was introduced, replacing the Gleason Score.}
				\label{table:tableGleason}
			\end{table}
			
\section*{Materials and methods} \label{Method}

		\subsection*{Neural network model}
			A convolutional neural network (CNN) was used to segment cribriform growth patterns in prostate biopsies stained with hematoxylin and eosin (H\&E). The network took as input $x(i)$ an 1024x1024 pixels RGB colored biopsy region, with $i$ indexing a particular pixel. The pixel sizes were 0.92x0.92 \SI{}{\micro\metre^2}/pixel in all cases (see \emph{Experiments} section).
			In order to better discriminate between cribriform and other G4 growth patterns, the network was trained to additionally detect other G3 and G4 tissue types. Accordingly, the output consisted of 7 probabilities $\hat{y}^l(i)$ where $l \in L$ being one of the following labels: non-labelled, G3, G4 fused, G4 ill-defined, G4 complex fused, G4 glomeruloid and G4 cribriform. The non-labelled class was included to represent the non-tissue background and any other non G3 and G4 tissue (healthy tissue, G5, mucinous, perineural growth...). 
			
			Henceforth, $y^l(i)$ represents the reference label, which consists of 7 masks, again with $i$ indexing a particular pixel. Each such reference label contained a 1 for one particular class and 0 for all other classes.
			 
			The Dice coefficient which quantifies positive overlap between label and prediction was used as loss function to be minimized during training of the network. With a batch of $P$ images, the loss was defined as:
			\begin{equation}\label{eq:dice}
			\mathcal{L}_{\mathrm{D}} = - \frac{1}{P} \sum\limits_{p=1}^{P} \left( \sum\limits_{l \in L} w_l \frac{2 \sum\limits_{i} y_p^{l}(i) \hat{y}_p^{l}(i)}{\sum\limits_{i}(y_p^{l}(i) + \hat{y}_p^{l}(i)) + \epsilon} \right)\end{equation}
			in which $w_l$ is a weight associated with tissue label $l$. Our goal was to detect the cribriform growth pattern and therefore we used for that tissue a higher weight. Note that the other labels were included in the loss function only to obtain a better convergence during the training of the neural network. Practically, a weight of 0.4 was applied for the cribriform label and 0.1 to all other labels (so that weights sum to 1).
			
			Details of the network architecture are depicted in Fig.~\ref{fig:unet_classification3}. The design was based on the following two criteria. First, we focused on coarse cribriform growth pattern localization. Therefore, the resolution of the segmented output did not have to be as high as the original 1024x1024 input resolution. Instead, the outputs were masks of 32x32 pixels, which was sufficient to segment the smallest relevant glands. Accordingly, the 1024x1024 reference masks were downsampled to 32x32 in order to match the output of the neural network using an average pooling with a 32x32 kernel. Second, we wanted to achieve fast convergence and simultaneously accurate training. To accomplish this we applied several architectural features that have been shown to enhance training efficiency: residual connections with every convolution block \cite{he2016deep}, 2x2 strided convolutions effectively learning to downsample \cite{springenberg2015striving} and squeeze-and-excitation blocks for adaptively weighting channels in the convolution blocks \cite{hu2018squeeze}. The output of the network was not post-processed afterwards. 
		
		\subsection*{Implementation} \label{Implementation}
			In order to have representative training data with each iteration of back-propagation, we made sure that a training batch of images always contained all the 7 possible labels. Accordingly, the batch size of the neural network was chosen to be the size of the label set $L = 7$. 

			Every network was initialized with weights randomly sampled from a uniform distribution \cite{glorot2010understanding} and trained during 60000 iterations with the stochastic gradient descent optimizer with a learning rate of 0.01, a decay of $5\times10^{-4}$ and a momentum of 0.99. No explicit stopping criterion was included. Instead a validation set was used to choose optimal weights according to a metric as specified in the \emph{Experiments} section.
			The method has been implemented in Python using Keras 2.2.4 and TensorFlow 1.12 libraries. We used the Titan Xp and GTX 1080 Ti GPU's from Nvidia Inc. to perform the experiments.
						
			To avoid over-training and add more variability in the training set, on-the-fly data augmentation was performed on each input patch. Input patches with their associated label masks were randomly vertically and/or horizontally flipped, translated in the range of $\pm10\%$ of the image size, rotated (around the image center) by maximally 5 degrees and scaled with a factor in the range of 0.9 to 1.1. In histopathology, the staining method with H\&E can result in different image contrasts. Therefore, after normalizing the RGB values (yielding values for each channel between 0 and 1) a random intensity shift was globally applied to each color channel of every image with a magnitude in the range of $\pm0.05$. Furthermore, the full range of the intensity (0 to 1) in each channel was also randomly rescaled in a linear fashion between minimum value $\in[0,0.1]$ and maximum value $\in[0.9,1]$.
						
\section*{Experiments}
		The proposed network was trained and tested on an annotated biopsy dataset. However, cribriform growth pattern detection by pathologists themselves is not a trivial task. Consequently, uncertainties in the annotations occur, which hinders the training of the network. Therefore we let the misclassified biopsy regions by the algorithm in the previous experiment be annotated for a second time by the same pathologist. We did this for a detailed assessment of the misclassifications of the neural network, but also to quantify the reproducibility of annotating the growth patterns. Moreover, the image dataset from an extensive inter-observer study \cite{kweldam2016gleason} was used to evaluate our network in comparison to the assessment by 23 pathologists.

		\subsection*{Ethics statement}
			The research and the analysis of prostate tissues was approved by the institutional Medical Ethics Committee (MEC) from Erasmus University Medical Center, Rotterdam, The Netherlands (MEC-2018-1614) and samples were used in accordance with the “Code for Proper Secondary Use of Human Tissue in The Netherlands” as developed by the Dutch Federation of Medical Scientific Societies (FMWV, version 2002, update 2011). The institutional Medical Ethics Committee (MEC) from Erasmus University Medical Center, Rotterdam, The Netherlands (MEC-2018-1614) stated that the study was not subject to the “Medical Research Involving Human Subjects Act” (WMO, Wet Medisch-wetenschappelijk Onderzoek) and so waived the informed consent procedure.

		\subsection*{Cribriform detection performance} \label{Cribriform_detection_evaluation}
			The CNN was first trained and tested on a dataset of prostate tissue images from 128 biopsies (one WSI per biopsy; one biopsy per patient) acquired by the department of Pathology of Erasmus University Medical Center, Rotterdam, The Netherlands. We selected only one WSI per patient in order to include data from as many patients as possible for an acceptable processing time. These data concerned clinical prostate biopsies from 2010 to 2016 with acinar adenocarcinoma cancer and a Gleason Score 6 or higher. From each patient, the biopsy with the most tumor volume was selected. 132 biopsies were digitized after which 4 of them were excluded due to severe artefacts and too little tumor tissue. The biopsies were stained with H\&E and digitized using a NanoZoomer digital slide scanner (Hamamatsu Photonics, Hamamatsu City, Japan). The resulting images had a resolution of \SI{0.23}{\micro\metre}/pixel. Two genitourinary pathologists sat together and annotated in consensus the different regions of each biopsy using the ASAP software \cite{asap}. Subsequently, a label $l \in L$ was assigned to each such region according to the updated standard classification \cite{epstein20162014,kweldam2019grading}. % as illustrated in Fig.~\ref{fig:annotation_example_biopt044}.
			The biopsies contained mainly G3 and G4 carcinoma. G5, mucinous differentiation, perineural growth and prostatic intraepithelial neoplasia were also present in respectively 6, 6, 1 and 1 biopsy slides. These rare tissues were assigned to the non-labelled group.
			
			To discard the background region from the samples, a thresholding procedure was applied to the optical density transformation of the RGB channels \cite{litjens2016deep}:
			\begin{equation}\mathrm{OD}_c = -\mathrm{log}_{10}\frac{I_c}{I_\mathrm{max}}\end{equation}
			where $\mathrm{OD}_c$ is the optical density of the channel $c$, $I_c$ is the initial intensity and $I_\mathrm{max}$ is the maximum intensity measured for the concerned channel. We found that the background was easily identifiable by $ OD_c < 0.12$ in any of the channels.

			Subsequently, each biopsy was downsampled to a resolution of \SI{0.92}{\micro\metre}/pixel (i.e. by a factor 4 from the acquisition resolution) and subdivided in half-overlapping patches of 1024x1024 pixels (thus, taking steps of 512 pixels). Patches with more than 99.5\% of background were discarded. For training, all remaining patches from each biopsy in the training set were shuffled and fed to the CNN while making sure that all classes were present (see \emph{Implementation} section).%ref{Implementation}).
			During testing, all the patches of a test biopsy image were inferred by the CNN. Subsequently, to reassemble the full biopsy segmentation, only the 512x512 region in the center of each patch was kept in order to avoid overlapping segmentations. Also, we expected that the center part would yield the best classification accuracy as there is more context around it compared to the periphery of the patch.
			
			An 8-fold cross-validation training scheme was applied. The 128 prostate biopsy images were therefore distributed over 8 groups (see Table~\ref{table:tableDistrib}). As detection of cribriform growth pattern is the focus of this study, the biopsies were first partitioned such that the cribriform annotations were uniformly distributed. Subsequently, the remaining biopsies were split up in a way to yield an approximately equal distribution of labels. This was done by means of the bin packing algorithm \cite{binpacking}.
			
			In each fold, 6 groups of images were used to train the network. Furthermore, one group was taken as a validation set to select the optimal neural network weights from all the weights saved after each epoch of training. The metric $V$ to do so was a combination of the Dice function $\mathcal{L}_{\mathrm{D}}$ (eq.\ref{eq:dice}) and also the specificity in order to minimize false positives:

			\begin{equation}V = \alpha \mathcal{L}_{\mathrm{D}} + (1 - \alpha) \mathcal{L}_{\mathrm{S}}\end{equation}
			where $\alpha\in[0,1]$ is a weight factor that balances the two terms; $\mathcal{L}_{\mathrm{S}}$ is the negative average pixel specificity of a batch of $P$ images:

			\begin{equation}
			\mathcal{L}_{\mathrm{S}} = - \frac{1}{P} \sum\limits_{p=1}^{P} \left( \sum\limits_{l \in L} w_l \frac{\sum\limits_{i} (1-y_p^{l}(i))(1-\hat{y}_p^{l}(i))}{\sum\limits_{i}(1-\hat{y}_p^{l}(i)) + \epsilon} \right)\end{equation}
			
			Within each fold the network was trained four times during which patches were randomly shuffled to yield an altered training order. Furthermore, four weights were applied in the validation metric $V$: $\alpha \in [0.2,0.3,0.4,1]$. Empirically, we found that any $\alpha > 0.4$  yielded the same weights as the Dice-only metric $\mathcal{L}_{\mathrm{D}}$ ( $\alpha = 1$). In total 16 (=4*4) networks per fold were trained. In addition, an ensemble classifier was defined as the arithmetic mean of the predictions of the 16 networks.
			
			The remaining group (of 8) served as the test set to evaluate the performance of the 16 networks as well as the ensemble network.
			
			From the predictions, i.e. the probability maps for each biopsy, the performance of the cribriform detection was assessed with receiver operating characteristic (ROC) and free-response receiver operating characteristic (FROC) analyses.
			To do so, cut-offs on the cribriform probability were varied to select the cribriform pixels. Thereafter, for each cut-off, neighboring cribriform pixels were taken together to form cribriform regions.
			The cribriform regions were analyzed in two ways: biopsy-wise and annotation-wise. The biopsy-wise analysis considered a biopsy positive for \emph{reference} purposes if there was at least one annotation by the pathologists labelled as cribriform. Similarly, the \emph{prediction} of a biopsy was considered positive if there were at least one predicted cribriform pixel in it. Complementary, the annotation-wise analysis considered an annotated (\emph{reference}) cribriform region correctly detected if and only if at least one pixel of a \emph{predicted} cribriform region overlapped with it. 
			We performed both types of analyses while also studying the effect of only taking into account predicted regions with a cumulative pixel area larger than \SI{0.0150}{\milli\metre\squared}. This size was chosen since the smallest cribriform region in the annotations was \SI{0.0155}{\milli\metre\squared}.
			
			\begin{table}[ht]
				\centering
				\begin{tabular}{|a|c|c|c|c|c|c|c|c|c|}
					\hline
					\textbf{Groups} & \cellcolor[gray]{0.75}\textbf{1} & \cellcolor[gray]{0.75}\textbf{2} & \cellcolor[gray]{0.75}\textbf{3} & \cellcolor[gray]{0.75}\textbf{4} & \cellcolor[gray]{0.75}\textbf{5} & \cellcolor[gray]{0.75}\textbf{6} & \cellcolor[gray]{0.75}\textbf{7} & \cellcolor[gray]{0.75}\textbf{8} & \cellcolor[gray]{0.75}\textbf{Total}\\
					\hline
					\textbf{Number of biopsies} & 13 & 15 & 16 & 17 & 17 & 18 & 18 & 14 & 128\\
					\hline
					\textbf{G3} & 124/13 & 121/14 & 106/14 & 124/16 & 78/16 & 143/18 & 132/15 & 148/14 & 976/120\\
					\hline
					\textbf{G4 fused} & 24/7 & 9/5 & 22/8 & 20/4 & 26/9 & 27/8 & 48/12 & 21/5 & 197/58\\
					\hline
					\textbf{G4 ill-defined} & 57/6 & 75/11 & 44/6 & 51/11 & 51/11 & 41/11 & 51/9 & 71/9 & 441/74\\
					\hline
					\textbf{G4 complex fused} & 6/2 & 1/1 & 4/2 & 1/1 & 9/2 & 4/2 & 3/1 & 0/0 & 28/11\\
					\hline
					\textbf{G4 glomeruloid} & 15/1 & 1/1 & 54/7 & 14/2 & 34/4 & 32/4 & 27/7 & 6/3 & 183/29\\
					\hline
					\textbf{G4 cribriform} & 20/5 & 20/5 & 20/5 & 20/6 & 20/6 & 20/6 & 20/6 & 21/6 & 161/45\\
					\hline
				\end{tabular}
				\caption{Number of instances, i.e. regions/biopsies distributed over cross validation folds.}
				\label{table:tableDistrib}
			\end{table}
			
		\subsection*{Re-evaluation study} \label{Intra_observer_study}
			To increase our insight into wrong classifications of the network, the pathologists who made the initial annotations re-evaluated the false positive and false negative detected regions. While doing so, they were not informed of the classification outcome of the network nor of their own original annotation. The re-evaluation has been done more than one year after the initial annotations which we think is sufficient time for the pathologists to not recall their previous labelings. From each false negative and false positive region a 512x512 patch was extracted surrounding the center of gravity of the region. Practically, this provided sufficient context for the pathologists to classify glands. For each such patch, the pathologists only re-evaluated the center. As with the original annotations, the same 7 labels could be assigned.
			Simultaneously, a confidence level had to be indicated on a scale from 0 (meaning undecided), to 4 (highly confident). Furthermore, if the pathologist was in doubt about the growth pattern, secondary labels could be registered. The outcomes were summarized in a confusion table.

		\subsection*{Inter-observer study}  \label{inter_observer_evaluation}
			The performance of our method in relation to the inter-observer variability of the gland pattern annotations was evaluated based on the data from the inter-observer study performed by Kweldam et al. \cite{kweldam2016gleason}. This dataset contains 60 prostate histopathology images extracted after radical prostatectomies and was classified by 23 genitourinary pathologists. Kweldam et al. \cite{kweldam2016gleason} aimed to include 10 images classified as G3, 40 as G4 (10 per growth pattern) and 10 as G5. These were selected by two pathologists not involved in the subsequent assessment by the 23 raters. The selected prostate images were acquired with a NanoZoomer digital slide scanner (Hamamatsu Photonics, Hamamatsu City, Japan).
			To avoid ambiguity during the annotation, for each case a yellow line delineated the glands to be classified (Fig.~\ref{fig:io_examples}).
			
			We applied our neural network to this dataset and compared the cribriform detection with the assessments of the pathologists.
	
			We trained 8 versions of our neural network with the dataset described previously in the section \emph{Cribriform detection performance}.
			Following the distribution shown in table~\ref{table:tableDistrib}, each network was trained on 7 groups and one group was used to select 4 optimal neural network weights based on the validation metric $V$ while applying different $\alpha \in [0.2,0.3,0.4,1]$. We repeated this 4 times and iterating accross the 8 groups to obtain in total 128 (=4*4*8) different neural networks. The ensemble of the 128 trained neural network was applied to each image of the inter-observer dataset.

			To do so the images of this dataset were resampled to yield the same resolution as the training data: \SI{0.92}{\micro\metre}/pixel. The resulting images were fed in patches of 1024x1024 pixels into our networks (as above), after which only the predicted output within the contoured regions was retained. 
			If at least one pixel in the output was predicted as cribriform with a probability superior to a particular cut-off (between 0 and 1), we considered that a cribriform growth pattern was detected. Cut-offs to be applied were chosen based on the FROC curve derived from the validation set used during the training.

	\section*{Results}
		\subsection*{Cribriform detection performance} \label{result_cribri_detection}
\begin{figure} % h
				\centering
				\includegraphics[width=0.45\linewidth]{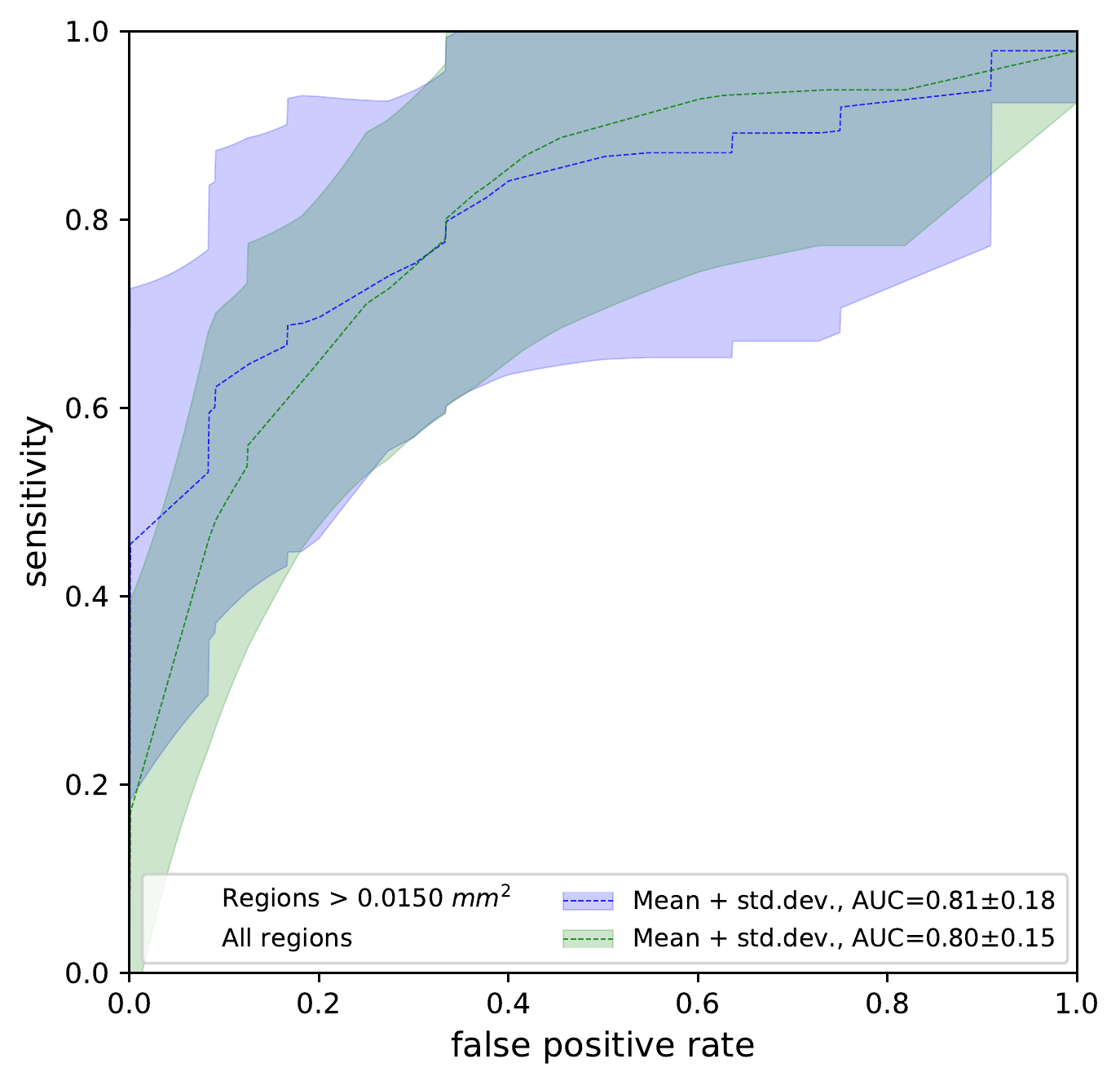}
				\includegraphics[width=0.45\linewidth]{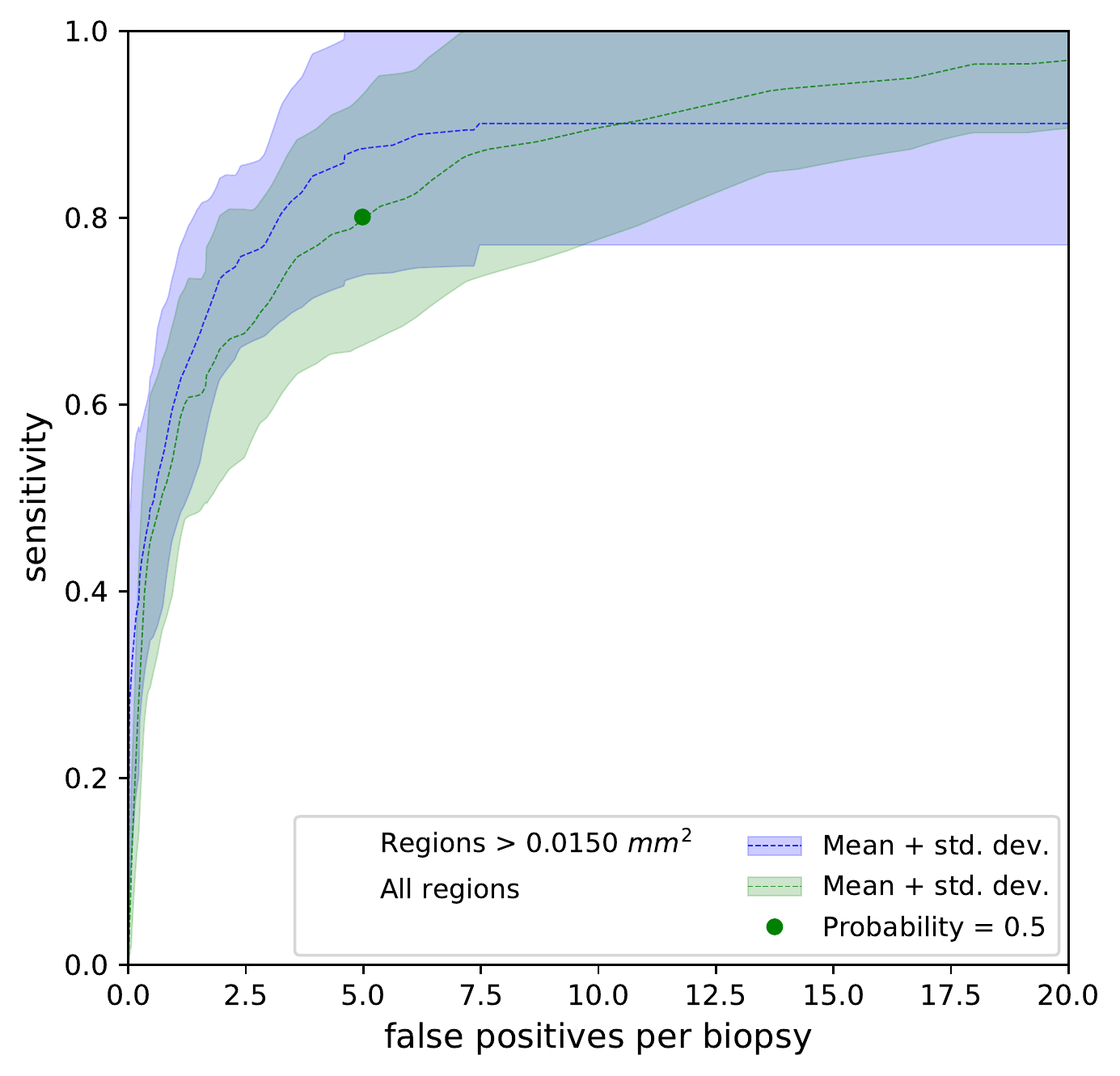}
				\caption{(left) Receiver operating characteristic (ROC) curves showing cribriform detection sensitivity per \emph{biopsy} as a function of the false positive rate (i.e. 1-specificity). Dashed lines represent mean ROC curves from the ensemble networks of the 8 folds; the associated shaded areas represent the corresponding standard deviations. The green curve concerns all predicted cribriform regions, the blue curve was based on predicted regions larger than \SI{0.0150}{\milli\metre\squared}.\\
				(right) Free-response receiver operating characteristic (FROC) curves of the cribriform detection sensitivity per \emph{annotation} as a function of the average number of false positive predictions per biopsy. The green dot indicates, for the mean FROC curve with all prediction regions, the cut-off probability of network output ($p=0.5$) conveniently chosen for the re-evaluation experiment (see \emph{Re-evaluation study} section).%ref{Re_evaluation_study}).
				}
				\label{fig:exp1_a}
			\end{figure}

			Fig.~\ref{fig:exp1_a} shows the ROC curves representing the cribriform detection sensitivity as a function of false positive rate per \emph{biopsy}. Complementary, it shows the FROC curves of the cribriform detection sensitivity per \emph{annotation} as a function of the mean number of false positive detections in biopsy. 
			
			In both figures, the dashed curves (and associated shaded areas) depict the mean performance (and corresponding standard deviation) of the ensemble network between the 8 folds. 
			The green curve collates results based on all cribriform prediction regions, whereas the blue curve considers only results of predicted regions larger than \SI{0.0150}{\milli\metre\squared}.
			
			The area under the curve (AUC) of the ensemble network in the ROC curve was on average 0.80 with all regions and 0.81 for regions larger than \SI{0.0150}{\milli\metre\squared}. 
			In order to compare the impact of the ensemble network with regards to the 16 individual networks, table~\ref{table:tableAUC} shows the AUC of the network ensemble and mean AUC of the 16 networks across the 8 folds.

\begin{figure} % h
				\centering
				\includegraphics[width=0.9\linewidth]{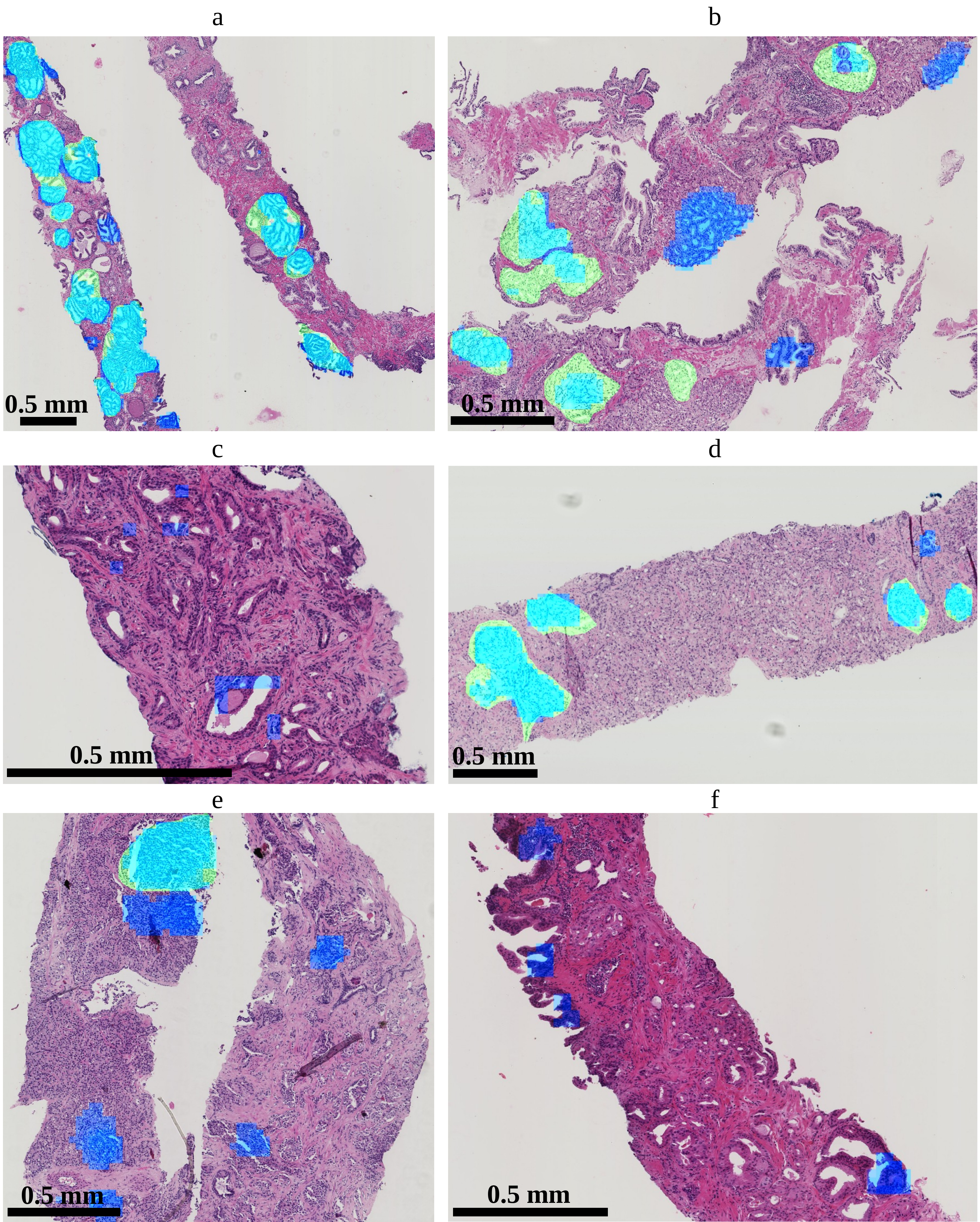}
				\caption{Biopsy slides with overlays showing predicted regions by the ensemble network as well as annotations (serving as reference). Light blue indicates true positive predictions; green are false negative regions; dark blue are false positive predictions.
				The scale of the images differs merely for illustration purposes.
				}
				\label{fig:exp1}
			\end{figure}

			Several representative example images with predictions and annotations are presented in Fig.~\ref{fig:exp1}.

			\sisetup{detect-all = true} % to use textbf with siunitx

			\begin{table}[ht]
				\centering
				\setlength\tabcolsep{3pt} % default value 6pt
				\begin{tabular}{|a|c|c|c|c|c|c|c|c|}
					\hline
					\textbf{Groups} & \cellcolor[gray]{0.75}\textbf{1} & \cellcolor[gray]{0.75}\textbf{2} & \cellcolor[gray]{0.75}\textbf{3} & \cellcolor[gray]{0.75}\textbf{4} & \cellcolor[gray]{0.75}\textbf{5} & \cellcolor[gray]{0.75}\textbf{6} & \cellcolor[gray]{0.75}\textbf{7} & \cellcolor[gray]{0.75}\textbf{8} \\
					\hline
					\textbf{Ensemble, all regions} & 0.80 & 0.84 & 0.89 & 0.85 & 0.41 & 0.79 & 0.92 & 0.91\\
					\hline
					\textbf{Mean, all regions} & 0.60$\pm$0.06 & 0.69$\pm$0.10 & 0.68$\pm$0.11 & 0.70$\pm$0.08 & 0.37$\pm$0.10 & 0.67$\pm$0.08 & 0.76$\pm$0.11 & 0.72$\pm$0.08\\
					\hline
					\textbf{Ensemble, regions > \SI{0.0150}{\milli\metre\squared}} & 0.84 & 0.81 & 0.92 & 0.87 & 0.36 & 0.86 & 0.92 & 0.93\\
					\hline
					\textbf{Mean, regions > \SI{0.0150}{\milli\metre\squared}} & 0.76$\pm$0.07 & 0.82$\pm$0.07 & 0.84$\pm$0.12 & 0.83$\pm$0.04 & 0.37$\pm$0.12 & 0.75$\pm$0.05 & 0.86$\pm$0.05 & 0.83$\pm$0.07\\
					\hline
				\end{tabular}
				\caption{AUC of the network ensemble and mean AUC of the 16 networks over cross validation folds.}
				\label{table:tableAUC}
			\end{table}

			\sisetup{detect-none = true} % we go back to normal state with siunitx

		\subsection*{Re-evaluation study}\label{Re_evaluation_study}
			In order to extract the false positive and negative regions of the ensemble network, we applied a cut-off to the cribriform prediction probability.
			We conveniently chose the cut-off to 0.5 (Fig.~\ref{fig:exp1_a}, right) in order to have a moderate amount of false positive regions to annotate, given the limited time the pathologists could allocate to this task.
			Applying the cut-off to the cribriform prediction probability yielded in total 632 false positive and 25 false negative cribriform region patches.

			During the re-evaluation, the pathologists gave `cribriform' as the first label to $9\% (=59/632)$ of the false positive patches. Furthermore, the pathologists indicated `cribriform' as the first or second label (the `doubtful' cases) to $20\%$ of false positive patches. 
			
			At the same time, upon re-evaluation the pathologists did not indicate `cribriform' as the first label in $44\% (=11/25) $ of the false negative cases. Furthermore, the pathologists did not indicate `cribriform' as the first nor as the second label in $16\%$ of those cases.
			
			For $71\% (=468/657$) of the wrongly classified patches (taking false positives and negatives together), the originally given label was identical to the first label during re-evaluation. Furthermore, in $48\%$ of the false positive patches and $36\%$ of the false negative patches, no second label was given.
			
			The median confidence level was 4 (highly confident) for patches with same labels during the initial annotation and the first annotation of the re-evaluation. The median confidence level was 2 for patches labelled differently as such. Furthermore, confidence level 4 was given to $51\%$ of the false positive patches and $28\%$ of the false negative patches during re-evaluation.
			
\begin{figure}[h] % h
				\centering
				\includegraphics[width=0.5\linewidth]{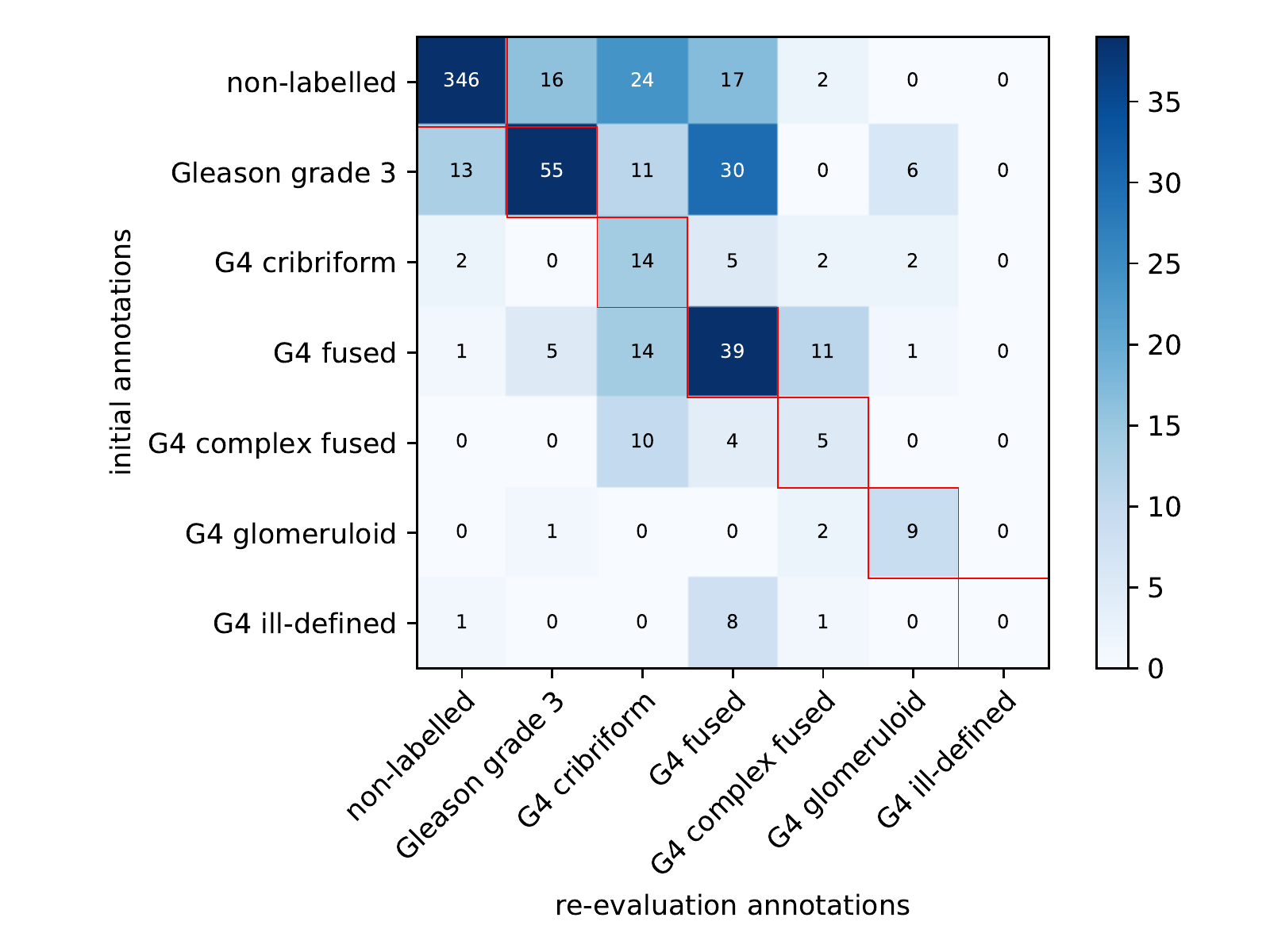}
				\includegraphics[width=0.43\linewidth]{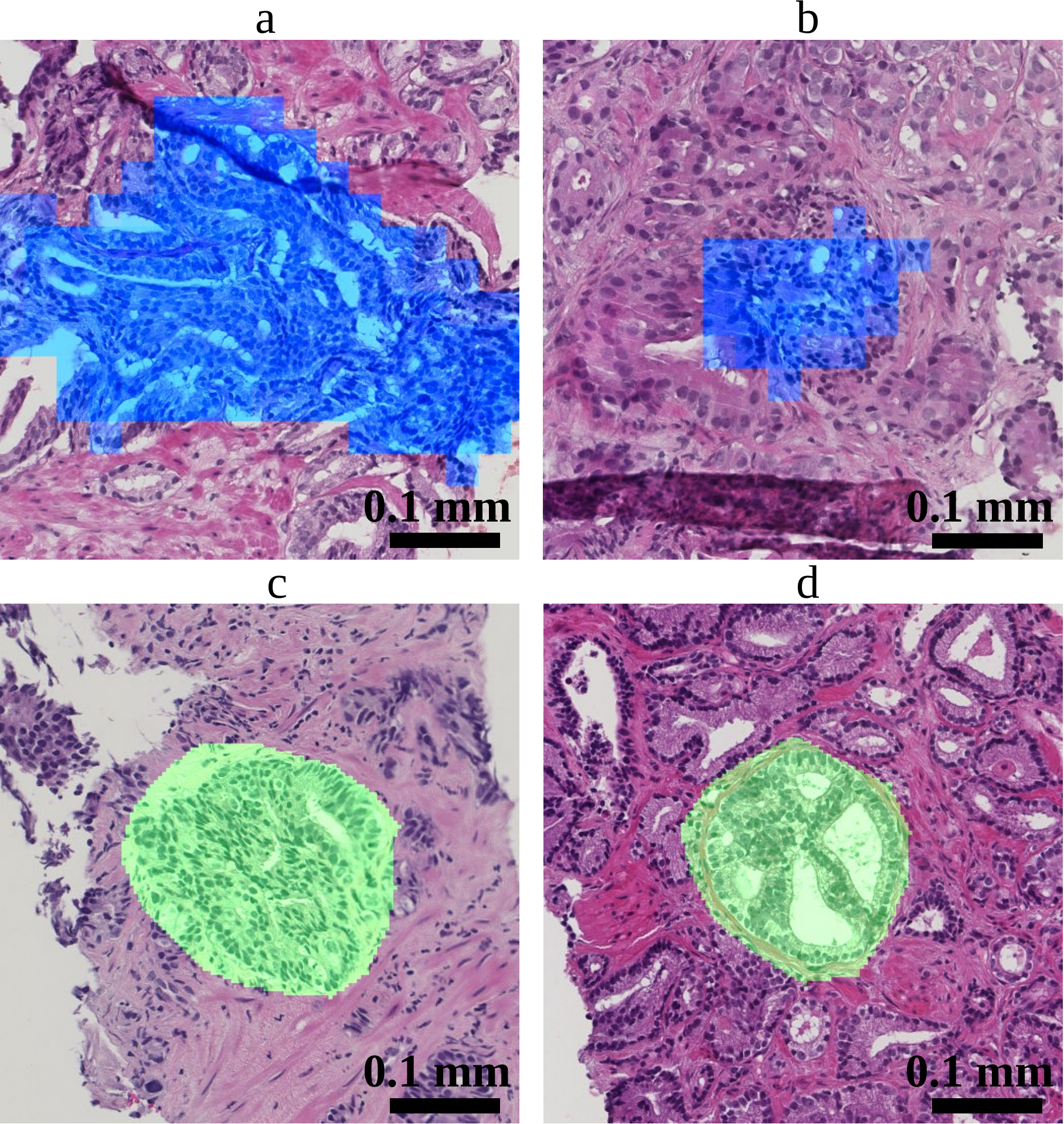}
				\caption{
				(left) Confusion matrix showing initial label versus first label during re-evaluation attributed by the pathologists to false positive and false negative patches.\\
				(right) Example false positive and false negative cases re-evaluated by the pathologists. Blue indicates a false positive detection and green indicates a false negative annotation.
				(a) False positive patch initially annotated as `fused' and given `cribriform' as the first label during re-evaluation.
				(b) False positive patch initially annotated as `fused' and given `complex fused' as the first label and `cribriform' as the second label during re-evaluation.
				(c) False negative patch initially annotated as `cribriform' and `fused' during re-evaluation.
				(d) False negative patch annotated as `cribriform' both during initial annotation and re-evaluation.\\
				Observe that the automatic segmentations (top) are coarse because the output of the neural network is 32 times smaller than the input; annotations (bottom) were manually drawn on the originals and are therefore smoother.
				}
				\label{fig:intra_examples}
			\end{figure}

			An overview of the initial annotations and first label during re-evaluation of the false positives and false negatives is contained in the confusion matrix in Fig.~\ref{fig:intra_examples}. Additionally, the figure shows examples of the false positive and negative cases including details on the annotations by the pathologist.
			
		\subsection*{Inter-observer study}
\begin{figure}[h] % h
				\centering
				\begin{minipage}{.43\linewidth}
					\includegraphics[width=1\textwidth]{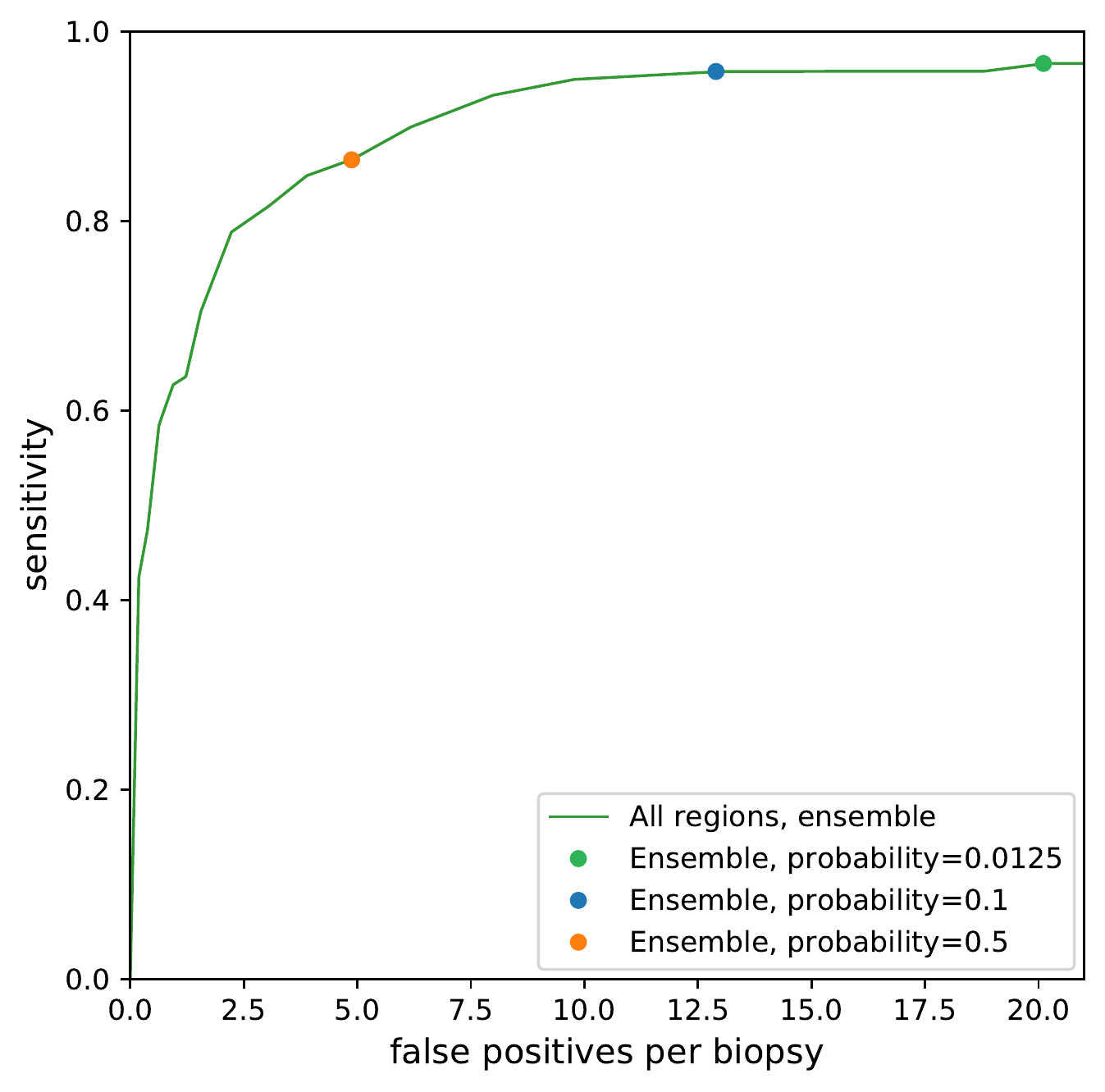}
				\end{minipage}
				\begin{minipage}{.51\linewidth}
					\includegraphics[width=1\textwidth]{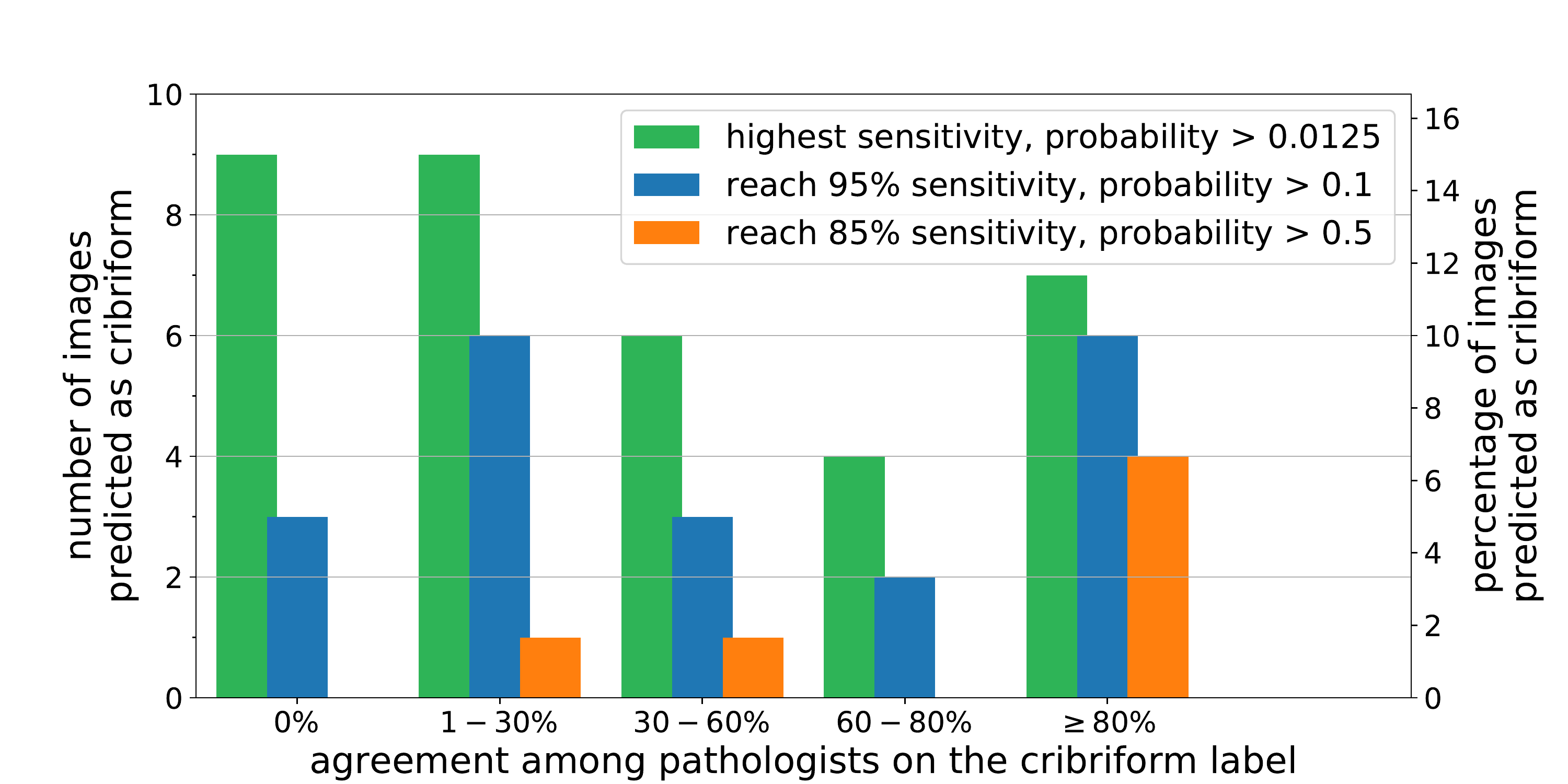}
					\includegraphics[width=1\textwidth]{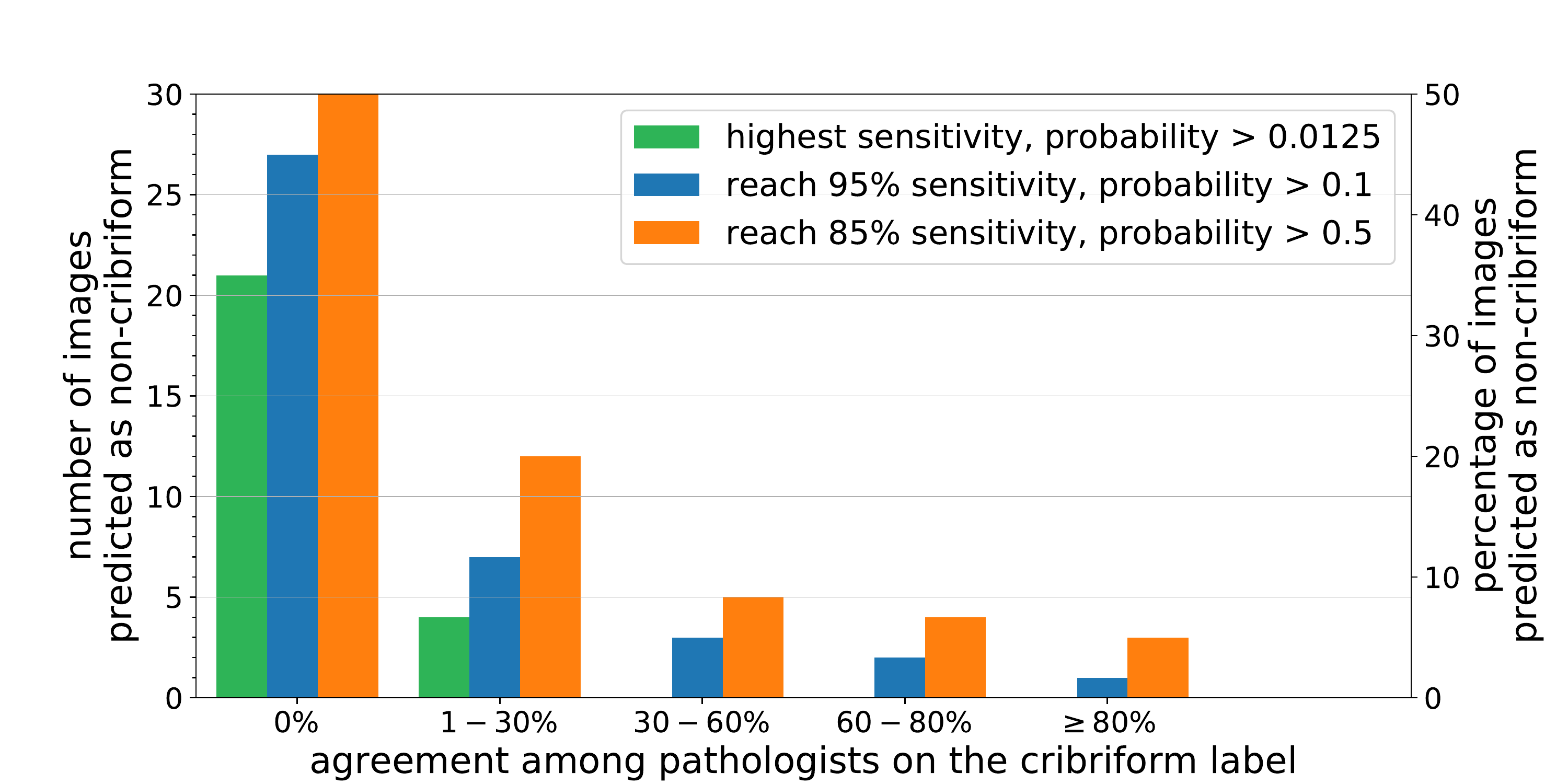}
				\end{minipage}
				\caption{(left) Free-response receiver operating characteristic (FROC) curve of the cribriform detection sensitivity per \emph{annotation} as function of the average number of false positive predictions per biopsy. This curve was resulted from the evaluation set using the ensemble of neural networks trained for the inter-observer study. The green dot corresponds with a probability cut-off at 0.0125, at which point the highest sensitivity is attained. The blue and orange dots correspond to probability cut-off at 0.1 and 0.5 yielding $95\%$ and $85\%$ sensitivity, respectively.\\
				(top-right) Numbers of images predicted as \emph{cribriform} by the neural network as a function of the percentage of pathologists labeling these images as cribriform.\\
				(bottom-right) Numbers of images \emph{not} predicted as cribriform by the neural network as a function of the percentage of pathologists labeling these images as cribriform.
				}
				\label{fig:100_196_203_small_cribriform_slide_froc_final}
			\end{figure}

			Fig.~\ref{fig:100_196_203_small_cribriform_slide_froc_final} shows the FROC curve of the ensemble network generated based on the evaluation set of the training data by varying the probability threshold.
			In order to compare the performance of the ensemble network to the assessments by the 23 pathologist we applied three probability thresholds/cut-offs: (1) 0.0125 corresponding to the highest attained sensitivity on the evaluation set; (2) 0.1 at which level $95\%$ sensitivity is reached; (3) 0.5 at which level $85\%$ sensitivity is reached.

			Top-right and bottom-right charts in Fig.~\ref{fig:100_196_203_small_cribriform_slide_froc_final} show the number of  regions predicted as \emph{cribriform} and \emph{not cribriform} respectively by the neural network as a function of the percentage of pathologists annotating these images as cribriform.
			
			Observe that there are 9, 3 and 0 regions labelled as cribriform by the network applying the thresholds at 0.0125, 0.1 and 0.5, respectively, which no pathologist annotated as cribriform.
			These could be considered as false positive cases.
			Furthermore, the 21, 27 and 30 regions with the same thresholds labelled as not-cribriform by the network nor labelled as cribriform by any pathologist could be considered true negatives.
			The bottom-right chart in Fig.~\ref{fig:100_196_203_small_cribriform_slide_froc_final} demonstrates that with the cut-off at 0.0125 \emph{all} the images annotated as cribriform by more than $30\%$ ($\geq$ 7/23) of the pathologists are predicted as cribriform by the neural network.
			Increasing the threshold to 0.1 and 0.5 leads to more regions not classified as cribriform and simultaneously less false positives.
			It may be noted that for 63\% (=19/30) of the cribriform cases, less than 60\% of the pathologists agreed regarding the labeling.

			The average of Cohen’s kappa coefficient across all paired pathologist labelings is 0.62. The average of Cohen’s kappa coefficient between our method and each pathologist is 0.29, 0.36 and 0.39 at cut-offs of 0.0125, 0.1 and 0.5, respectively.
			
\begin{figure} % h
				\centering
				\includegraphics[width=0.8\linewidth]{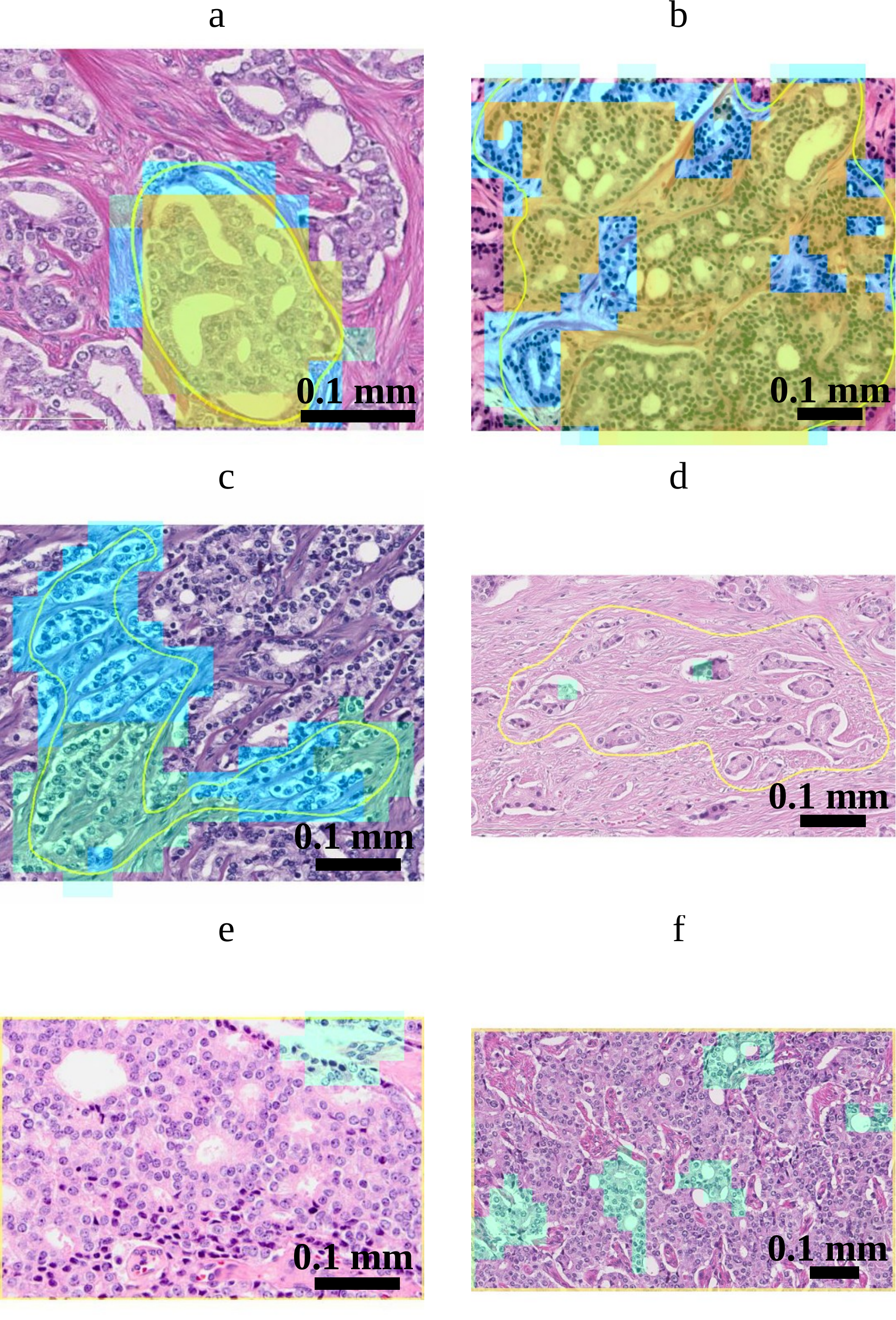}
				\caption{Cribriform growth pattern detection on images from the dataset of Kweldam et al. \cite{kweldam2016gleason}. The yellow contours delineate the regions that were assessed by the pathologists.
				Green, blue and orange colors correspond to detection of cribriform regions at probability thresholds of 0.0125, 0.1 and 0.5 respectively.
				(a),(b) Images annotated as cribriform by  22 and 23 pathologists, respectively. (c),(d) Images annotated principally as ill-defined and as G3, respectively (no cribriform annotation).
				(e),(f) Images annotated as cribriform by 22 and 8 pathologists, respectively. Image (f) was annotated as fused by 13 pathologists. Note that the yellow contours for images (e) and (f) is at the image border.}
				\label{fig:io_examples}
			\end{figure}

			Fig.~\ref{fig:io_examples} shows examples images for varying agreements between the pathologists on which the cribriform regions detected by the neural network are overlaid.
		
	\section*{Discussion}
		We proposed a method to automatically detect and localize G4 cribriform growth patterns in prostate biopsy images based on convolutional neural networks. In order to improve the detection of cribriform growth patterns the network was trained to detect other growth patterns (complex fused, glomeruloid, ...) as well. Furthermore, to enhance the stability of the prediction, an ensemble of networks was trained after which the average prediction was used.
		
		The ensemble networks focusing on regions larger than \SI{0.0150}{\milli\metre\squared} reached a mean area under the curve of 0.81 regarding detection of biopsy images harboring a cribriform region. The FROC analysis showed that achieving a sensitivity of 0.9 for regions larger than \SI{0.0150}{\milli\metre\squared} goes at the expense of on average 7.5 false positives per biopsy.
		
		The evaluation of the 16 individual neural networks show marked variation in AUC value. This could be an indication that the training data order (stochastic variation) influences the performance of the network. Our solution to cope with this was to apply the ensemble of neural networks. An alternative solution could be hard negative mining by prioritizing 'difficult' regions during training. In this way, the training process would more frequently present patches with large classification discrepancy across training iterations and thus would reduce the influence of the training data order. On the other hand, the AUC variation between the folds of the cross-validation has an even higher standard deviation than the stochastic variation on the ROC and FROC curves. Of course, yet another solution would be to acquire a larger training set, to take into account the heterogeneity of the data in a better way.

		During the evaluation of the method, the variations among pathologists regarding cribriform growth pattern recognition was also studied. In particular, false positive and false negative patches from the ensemble classifier were re-evaluated by the same pathologists.
		
		The evaluation of the method with the original annotations showed that sensitive cribriform region detection can be done, but at the expense of a high number of false positives. However, the re-evaluation study demonstrated that up to 20\% of the false positive detections could actually be cribriform regions. Concurrently, it showed that up to 44\% of the false negatives might not be cribriform regions.
		
		We also tested the ensemble network on a dataset annotated by 23 pathologists to put its performance into perspective regarding inter-observer variability. In 63\% of the cases, less than 60\% of the pathologists agreed regarding the cribriform labeling. Using a probability cut-off at 0.0125 (corresponding to the highest sensitivity in the training set) \emph{all} images annotated as cribriform by at least $30\%$ of the pathologists were also predicted as cribriform by the neural network.
		In other words, the network is rather conservative in classifying regions as cribriform even with a low percentage of agreement, which is opposed to detecting only regions for which there is large agreement. This is also reflected by the Cohen’s kappa which showed higher agreement amongst pathologist than between our method and the pathologists.
		As such, with the cut-off at 0.0125, the network is more inclusive than the consensus of pathologists. This could be clinically relevant as preferably no potentially cribriform region should be missed by the automated detection algorithm at this stage.

		Some fused and tangentially sectioned glands were falsely labelled as cribriform. In practice, biopsies are cut at three or more heights giving additional information to the pathologist, while we only used one level in the current study. We believe that the performance for recognition of cribriform architecture or grading in general can be improved if information of different levels of the same biopsy specimen can be registered and integrated in the future. Furthermore, due to its relatively large size, cribriform architecture may not be visualized in its entirety in biopsies, which is different from the situation in operation specimens from radical prostatectomies. Therefore, optimal training sets for cribriform pattern should be slightly different for biopsy and prostatectomy specimens.

		There were typical false positive cribriform detections. First, as malignant glands are not properly attached to the surrounding stroma, tearing may happen during tissue processing. Figs~\ref{fig:io_examples}.c and~\ref{fig:io_examples}.d show resulting retraction and slit-like artifacts surrounded by cells with clear cytoplasm. The resulting background could be mistaken for cribriform lumina by the network. Second, in Figs~\ref{fig:exp1}.c and~\ref{fig:exp1}.f, some regions in non-labelled tissue have complex anastomosing glands, not meeting the criteria for cribriform growth. Finally, in the latter figure, the background seems confused by the network for cribrifrom lumina. Since these areas are but a small part of the total area of non-labelled tissue it could be that the network might not have seen sufficient examples of such patterns during the training.

		The latter observation signifies the importance of diversity in the training data also for our application. Inclusion of healthy tissue simultaneously with rare tissues such as G5, mucinous, and perineural structures are indispensable in the training set. For similar reasons, an important direction for future work could be to particularly focus on multi-center data. Also, annotations from multiple pathologists might help to build a more detailed probabilistic model and cover the variability from large consensus to large ambiguity. Previously, such approaches were proposed by Nir et al. \cite{nir2018automatic} for Gleason grading and by Kohl et al. \cite{NIPS2018_7928} for segmentation of lung abnormalities. %with annotations from multiple radiologists.

		A limitation of our study is that a re-evaluation by the pathologists of true positives and true negatives cases is lacking. While our re-evaluation analysis indicates that some of the false positives and false negatives cases may not necessarily be false, it is likely that reassessing all the samples would also turn some true positives and true negatives patches into false positives and false negatives, respectively. Furthermore, as it is stated in Kweldam et al. \cite{kweldam2016gleason}, the dataset from the inter-observer study has been deliberately chosen to be difficult which may lead to more disagreement between pathologists than with biopsy analyses during daily clinical practice. The performance of the neural network is also impacted by this as such uncommon cases were not present in the training.
		Also, despite its predictive value, no global consensus exists yet on the definition of cribriform architecture and its delineating features from potential mimickers. The pathologists who annotated our dataset have however shown statistically significant correlation of cribriform pattern with clinical outcome in large biopsy and operation specimens cohorts, clinically validating the criteria used in this study \cite{vanprostate,hollemans2019large,kweldam2015cribriform}.

	\section*{Conclusion}
		We proposed a convolutional neural network to automatically detect and localize cribriform growth patterns in prostate biopsy images. The ensemble network reached a mean area under the curve of up to 0.81 for detection of biopsies harboring cribriform tissue. This result must be valued taking into account the large disagreement among pathologists.
		The network is showing rather conservative performance: cases were detected as cribriform even when just a limited number of pathologists labelled them as such. The method could be clinically useful by serving as a sanity check, to avoid missing cribriform patterns.
		
	\section*{Data availability}
		Data from the inter-observer study performed by Kweldam et al. \cite{kweldam2016gleason} are available as an appendix at https://doi.org/10.1111/his.12976.

\bibliography{cribriform}

% \noindent LaTeX formats citations and references automatically using the bibliography records in your .bib file, which you can edit via the project menu. Use the cite command for an inline citation, e.g.  \cite{Hao:gidmaps:2014}.

% For data citations of datasets uploaded to e.g. \emph{figshare}, please use the \verb|howpublished| option in the bib entry to specify the platform and the link, as in the \verb|Hao:gidmaps:2014| example in the sample bibliography file.

\section*{Acknowledgements}
	This work was funded by the Netherlands Organization for Scientific Research (NWO), project TNW 15173.
	We thank NVIDIA Corporation for the donation of a Titan Xp via the GPU Grant Program.
	Muhammad Arif and Jifke Veenland of the Erasmus University Medical Center are acknowledged for fruitful discussions and advice.

\section*{Author contributions statement}
P.A. designed and performed the algorithms and experiments, analyzed the results and wrote the manuscript.
E.H. and G.J.L.H.L. performed the data collection, grading of the biopsy set and brought medical expertise.
C.F.K. performed the data collection of the inter-observer set and brought medical expertise.
S.S. and F.V. supervised the work and wrote the manuscript.
All authors reviewed and revised the manuscript.

\section*{Additional information}
\textbf{Competing interests} The authors declare no conflict of interest.

\end{document}